\documentclass[aps,prl,twocolumn,superscriptaddress,floatfix, 10pt]{revtex4-2}
	
	\usepackage{amsmath,amssymb,bm, physics}
	\usepackage{graphicx}
	\usepackage{hyperref}
	\usepackage{xcolor}
	\usepackage[normalem]{ulem}
	\usepackage{xurl}
	
\begin{document}
		
		\title{Chiral Vacuum Engineering of Quantum Hall Matter}
		\author{Kiran M. Kulkarni}
		\email{kirankulkarni@rice.edu}
		\affiliation{\rm Applied Physics Graduate Program, Smalley-Curl Institute, Rice University, Houston, TX, USA.}
		\affiliation{\rm Department of Electrical and Computer Engineering, Rice University, Houston, TX, USA.}
		
		\author{Andrey Baydin}
		\email{baydin@rice.edu}
		\affiliation{\rm Department of Electrical and Computer Engineering, Rice University, Houston, TX, USA.}
		\affiliation{\rm Smalley--Curl Institute, Rice University, Houston, TX, USA.}
		
		\author{Motoaki Bamba}
		\affiliation{\rm Department of Physics, Graduate School of Engineering Science, Yokohama National University, Hodogaya-ku, Yokohama, Japan.}
		\affiliation{\rm Institute for Multidisciplinary Sciences, Yokohama National University, Hodogaya-ku, Yokohama, Japan}
		
		
		\author{Junichiro Kono}
		\email{kono@rice.edu}
		\affiliation{\rm Department of Electrical and Computer Engineering, Rice University, Houston, TX, USA.}
		\affiliation{\rm Smalley--Curl Institute, Rice University, Houston, TX, USA.}
		\affiliation{\rm Department of Physics and Astronomy, Rice University, Houston, TX, USA.}
		\affiliation{\rm Department of Materials Science and NanoEngineering, Rice University, Houston, TX, USA.}
		
		\author{Ceren B. Dag}
		\email{cbdag@iu.edu}
		\affiliation{\rm Department of Physics, Indiana University, Bloomington, IN, USA.}
		\affiliation{\rm Department of Physics, Harvard University, Cambridge, MA, USA.}
		
		\author{Vasil Rokaj}
		\email{vasil.rokaj@villanova.edu}
		\affiliation{\rm Department of Physics, Villanova University, Villanova, PA, USA.}
		
		\vspace{5em}
		\date{\today}
		
		\begin{abstract}
			We reveal that a chiral cavity distinguishes the resonant and antiresonant interactions between a two-dimensional electron gas and the cavity vacuum. For co-rotating helicities, the modes anticross, exhibiting vacuum Rabi splitting, and retain a bare vacuum ground state. For counter-rotating helicities, the modes cross, acquiring vacuum-mediated energy shifts and a two-mode squeezed ground state. We trace the origin of these observations to the SU(2) and SU(1,1) symmetries of the respective Hamiltonians and present their spectroscopic signatures in chiral photonic crystal cavities. We further predict that cavity chirality produces a handedness-dependent correction to Hall resistivity at finite polariton broadening, while such a correction is not present in an achiral cavity. For the longitudinal resistivity, we find an asymmetric response of the Shubnikov--de Haas oscillations for the different helicities. Finally, we present an experimentally accessible route for detecting the squeezed ground state through current fluctuations. Parameters based on a recently demonstrated terahertz chiral cavity place these effects within experimental reach.
		\end{abstract}
		
		\maketitle
		
		Recent advances in cavity engineering of quantum materials~\cite{SchlawinEtAl2022APR, QinEtAl2024PR, LuEtAl2025AOPA, baydin2025perspective, bretscher2026fluctuation} have revealed cavity-induced modifications of superconductivity ~\cite{ThomasEtAl2025JCP, KerenEtAl2026N}, magneto-transport in quantum Hall systems~\cite{ParaviciniEtAl2019NP, AppuglieseEtAl2022S, EnknerEtAl2025N}, and magnetism~\cite{ThomasEtAl2021NL}. In parallel, an increasing number of theoretical predictions have explored the effects of cavity vacuum fields on material properties, with proposals ranging from cavity-enhanced superconducting pairing~\cite{SentefEtAl2018SA, SchlawinEtAl2019PRL, CurtisEtAl2019PRL} to vacuum-induced topological phases~\cite{WangEtAl2019PRB, DagRokaj2024PRB}. 
			Controlling matter through the symmetry of cavity vacuum fields remains an open question. In this regard, several chiral cavities have been proposed and realized~\cite{silverman1994interferometric,bao2020chirality,voronin2022single}, with some suitable for reaching the ultrastrong light--matter coupling (USC) regime~\cite{kulkarniRealizationChiralPhotoniccrystal2026,tayTerahertzChiralPhotoniccrystal2025a}. 
		
		Here, we investigate chiral light--matter interactions in a two-dimensional electron gas (2DEG) perpendicular to a static magnetic field and coupled to a circularly polarized cavity mode that breaks time-reversal symmetry (TRS), motivated by the experimental realization of a chiral photonic crystal cavity~\cite{kulkarniRealizationChiralPhotoniccrystal2026}. We provide the complete analytical solution for the 2DEG coupled to the chiral vacuum field for both helicity sectors at arbitrary coupling, including the diamagnetic interaction. Thus, our model is applicable to all regimes of light--matter interaction and goes beyond previous perturbative treatments~\cite{yang2025quantum}. In comparison to studies of Landau polaritons in linearly polarized and achiral cavities~\cite{rokajWeakened, RokajResistivity2026, BartoloCiuti2018,zhangCollectiveNonperturbativeCoupling2016,liVacuumBlochSiegert2018}, where the co- and counter-rotating interactions are inseparable, our chiral-cavity setup~\cite{tayTerahertzChiralPhotoniccrystal2025a, kulkarniRealizationChiralPhotoniccrystal2026} allows us to isolate and study the effects of these interactions separately. 
		
		In our work, we present theoretical predictions for experimentally accessible signatures of chiral vacuum fields in a 2DEG. When the vacuum polarization co-rotates with the 2DEG electrons, the cyclotron-resonance-active (CRA) hybrid modes avoid each other. When it counter-rotates, the chiral-vacuum-dressed modes of the cyclotron-resonance-inactive (CRI) interaction instead cross, while acquiring vacuum-mediated energy shifts, an enhanced Bloch--Siegert shift, and a squeezed ground state. The spectroscopic signatures of the CRA and CRI modes are distinct from their achiral counterparts. The DC transport simulations at zero temperature reveal that the chiral electromagnetic environment around the 2DEG modifies the Hall resistivity through the polariton broadening and the light--matter coupling strength. This is in contrast to the case of an achiral cavity field, where the Hall resistivity is entirely immune to polariton broadening and light--matter coupling at zero temperature~\cite{RokajResistivity2026}. In the counter-rotating configuration, the transverse current fluctuations fall below the bare vacuum limit, providing a physical observable of the squeezed ground state~\cite{ciutiQuantumVacuumProperties2005, MakiharaEtAl2021NC}. 
		
\begin{figure*}
			\centering
			\includegraphics[width=0.85\textwidth]{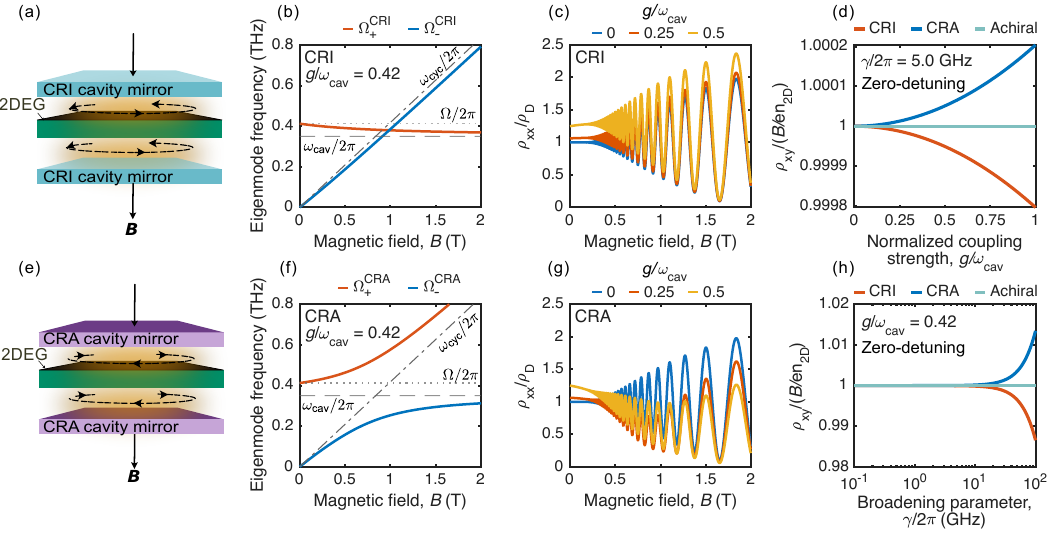}
			\caption{
				(a),(e)~A 2DEG in a chiral cavity confining right- and left-circularly polarized light, respectively.
				(b)~Eigenspectrum of the CRI Hamiltonian: the two chiral-vacuum-dressed modes cross, with the cavity-like branch renormalized to $\Omega = \omega_\mathrm{cav} + g^2/\omega_\mathrm{cav}$ by the diamagnetic term and the cyclotron-like branch by the vacuum Bloch--Siegert shift.
				(f)~Eigenspectrum of the CRA Hamiltonian: two polariton branches with avoided crossing.
				(c), (g)~Longitudinal resistivity vs.\ magnetic field $B$ for various coupling strengths, for the CRI and CRA systems, respectively, normalized by the Drude resistivity $\rho_\mathrm{D}= m^*\gamma/(n_\mathrm{2D}e^2)$, where $n_\mathrm{2D}=2\times10^{11}\,\mathrm{cm}^{-2}$. We observe Shubnikov de Haas (SdH) oscillations as a function of $B$. The SdH oscillations are modeled through a field-dependent broadening parameter $\gamma_r(B)$. The CRA system exhibits damped SdH oscillations, due to strong light--matter hybridization, whereas the CRI system shows undamped SdH oscillations, owing to its chiral-vacuum-dressed modes. 
				(d),(h)~Hall resistivity $\rho_{xy}$, normalized by the classical Hall resistivity $(B/e\,n_{\rm 2D})$, at fixed $g/\omega_\mathrm{cav}=0.42$ and $\gamma/2\pi=5$\,GHz. The achiral cavity case shows no cavity-induced correction, coinciding with the bare 2DEG case. In the cases of chiral cavities, there is a handedness-dependent correction; see Eq.~\eqref{Eq: rho chiral}.
				Here, $\omega_{\rm cav}/2\pi = 0.35$\,THz, $g/\omega_\mathrm{cav} = 0.42$, $\Omega/2\pi= 0.412$\,THz, and $\gamma/2\pi = 5$\,GHz, fixed for panels (b),(d),(f),(h). Dashed line marks the cavity frequency, dotted marks the renormalized cavity frequency $\Omega$, dash-dot marks the cyclotron frequency.}
			\label{Fig: fig1}
		\end{figure*}

		\textit{Model---}We consider a 2DEG under a perpendicular magnetic field $\boldsymbol{B} = B \boldsymbol{e}_z$, which quantizes the electron energy into Landau levels with cyclotron frequency $\omega_{\rm cyc} = eB/m^*$ and filling factor $\nu = n_{\rm 2D} h / eB$, where $e$ is the electronic charge, $m^*$ is the effective electron mass, $n_{\rm 2D}$ is the electron sheet density, and $h$ is the Planck constant. The 2DEG is embedded in a chiral photonic cavity that supports a cavity mode of single-handedness at frequency $\omega_{\rm cav}$, as shown in Figs.~\ref{Fig: fig1}(a) and (e). 
		The interacting system is described by the Pauli--Fierz Hamiltonian~\cite{cohen-tannoudjiPhotonsAtomsIntroduction2004}
		\begin{equation}
			\label{Eq: H_full}
			\hat{\mathcal{H}} = \sum_{i=1}^{N}\frac{\bigl(\hat{\boldsymbol\pi}_i + e\hat{\boldsymbol{A}}_\mathrm{cav}\bigr)^2}{2m^*}
			+ \hbar\omega_\mathrm{cav}\left(\hat a_\mathrm{L/R}^\dagger\hat a_\mathrm{L/R} + \frac{1}{2}\right) + \hat V_\mathrm{ee},
		\end{equation}
		where $\hat{\boldsymbol\pi}_i \equiv i\hbar\boldsymbol\nabla_i + e\boldsymbol{A}_\mathrm{ext}(\boldsymbol{r}_i)$ is the kinetic momentum of the $i$-th electron under the external magnetic field, $\boldsymbol{A}_\mathrm{ext} = -B y\,\boldsymbol{e}_x$ is the Landau-gauge vector potential, the operators $\hat a_{\mathrm{L/R}}$ and $\hat a^\dagger_\mathrm{L/R}$ are the annihilation and creation operators of the left/right circularly polarized cavity mode, satisfying bosonic commutation relations $[\hat a_\mathrm{L/R}, \hat a^\dagger_\mathrm{L/R}] = 1 $, and $\hat V_\mathrm{ee}$ denotes the electron--electron interaction. The circularly polarized cavity field, assumed to be spatially uniform, is described by the vector potential $\hat{\boldsymbol{A}}_\mathrm{cav} = A_0\bigl(\boldsymbol{e}_\mathrm{L/R}\,\hat a_\mathrm{L/R} + \boldsymbol{e}_\mathrm{L/R}^*\,\hat a^\dagger_\mathrm{L/R}\bigr)$ with left/right-circular polarization vectors $\boldsymbol{e}_\mathrm{L/R} = (\boldsymbol{e}_x \mp i \boldsymbol{e}_y)/\sqrt{2}$, field amplitude $A_0 = \sqrt{\hbar/(2\varepsilon_0\varepsilon_r\mathcal{V}_\mathrm{eff}\omega_\mathrm{cav})}$, $\varepsilon_r$ the background permittivity, and $\mathcal{V}_\mathrm{eff}$ the effective mode volume of the cavity.
		
		By Kohn's theorem~\cite{Kohn1961}, a spatially uniform electromagnetic field excites only the center-of-mass (CM) coordinate of the electrons, $\boldsymbol{R} = (X, Y)= \sum_{j=1}^N \boldsymbol{r}_j/\sqrt{N}$, while the electronic interactions $\hat V_\mathrm{ee}$ decouple from the field and the CM. This separation also extends to uniform quantized cavity fields~\cite{rokajWeakened}. The CM part of Eq.~(\ref{Eq: H_full}), up to a constant zero-point energy, can be written as 
		\begin{equation}
			\hat{\mathcal{H}}_\mathrm{CM} = \frac{1}{2m^*} (\hat{\boldsymbol{\Pi}} + e\sqrt{N}\hat{\boldsymbol{A}}_\mathrm{cav} )^2 + \hbar \omega_\mathrm{cav}\,\hat a^\dagger_\mathrm{L/R}\hat a_\mathrm{L/R}
		\end{equation}
		where $\hat{\boldsymbol{\Pi}} = i\hbar \nabla_{\mathbf{R}} + e\boldsymbol{A}_\mathrm{ext}(\boldsymbol{R})$ is the kinetic momentum of the CM. Expressing the CM cyclotron motion in terms of Landau ladder operators $\hat{b}$, $\hat{b}^\dagger$ satisfying $[\hat{b}, \hat{b}^\dagger] = 1$ (Supplemental Material- SM), $\hat{\mathcal{H}}_\mathrm{CM}$ reduces to one of two forms depending on the relative handedness of the cavity field and the cyclotron orbit.
		
		When the cavity field counter-rotates with the cyclotron motion (cyclotron-resonance-inactive, CRI), the Hamiltonian describing the antiresonant interaction is
		\begin{equation}
			\label{Eq: H_CRI}
			\hat{\mathcal{H}}_\mathrm{CRI} = \hbar\Omega\,\hat a_{\mathrm{R}}^\dagger\hat a_{\mathrm{R}}
			+ \hbar\omega_\mathrm{cyc}\,\hat b^\dagger\hat b
			- \hbar\bar g\bigl(\hat a_{\mathrm{R}}^\dagger\hat b^\dagger + \hat a_{\mathrm{R}}\,\hat b\bigr),
		\end{equation}
		where the collective coupling constant ($g$), vacuum Rabi frequency ($\bar g$), diamagnetic coefficient ($D$), and renormalized cavity frequency ($\Omega$) are 
		\begin{align}
			\label{Eq: couplings}
			g &= \sqrt{\frac{Ne^2}{2m^*\varepsilon_0\varepsilon_r\mathcal{V}_{\rm eff}}},\qquad
			\bar g = g\sqrt{\frac{\omega_{\rm cyc}}{\omega_{\rm cav}}}, \notag\\
			D &= \frac{g^2}{\omega_{\rm cav}} = \frac{\bar g^2}{\omega_{\rm cyc}},\qquad
			\Omega = \omega_{\rm cav} + D.
		\end{align}
		
		When the cavity field co-rotates with the cyclotron motion (CRA), the Hamiltonian becomes: 
		\begin{equation}
			\label{Eq: H_CRA}
			\hat{\mathcal{H}}_\mathrm{CRA} = \hbar\Omega\,\hat a_{\mathrm{L}}^\dagger\hat a_{\mathrm{L}}
			+ \hbar\omega_\mathrm{cyc}\,\hat b^\dagger\hat b
			- \hbar\bar g\bigl(\hat a_{\mathrm{L}}\,\hat b^\dagger + \hat a_{\mathrm{L}}^\dagger\hat b\bigr),
		\end{equation}
		containing only resonant interaction terms that describe the coherent exchange of quanta between light and matter modes. 
		
		\textit{Eigenspectrum and symmetries---} We diagonalize $\hat{\mathcal{H}}_\mathrm{CRI}$ using the Hopfield--Bogoliubov transform~\cite{hopfieldTheoryContributionExcitons1958a} and find the eigenmodes
		\begin{equation}
			\label{Eq: CRI eigs}
			\Omega_\pm^{\rm CRI} = \tfrac{1}{2}\left(\Sigma \pm (\Omega - \omega_{\rm cyc})\right); \; \Sigma \equiv \sqrt{{\left(\Omega + \omega_{\rm cyc}\right)}^2 - 4\bar g^2}.
		\end{equation}
		These \textit{chiral-vacuum-dressed} modes cross, and are shifted from the bare cavity and cyclotron frequencies by the diamagnetic and vacuum Bloch–Siegert terms, respectively (Fig.~\ref{Fig: fig1}b). Diagonalizing  $\hat{\mathcal{H}}_\mathrm{CRA}$ yields
		\begin{equation}
			\label{Eq: CRA eigs}
			\Omega_\pm^{\rm CRA} = \tfrac{1}{2}\left(\Omega + \omega_{\rm cyc} \pm \sqrt{{\left(\Omega - \omega_{\rm cyc}\right)}^2 + 4\bar g^2}\right).
		\end{equation}
		These modes exhibit avoided crossing at zero detuning ($\Delta = \Omega - \omega_{\rm cyc}$), with a vacuum Rabi splitting of $2\bar g = \Omega_+^{\rm CRA} - \Omega_-^{\rm CRA}$ between the upper ($+$) and lower ($-$) polariton modes; see Fig.~\ref{Fig: fig1}f. 
		
		The counter-rotating terms in $\hat{\mathcal{H}}_\mathrm{CRI}$ can be expressed in terms of the generators of an SU(1,1) algebra $\hat K_+ = \hat a_{\mathrm{R}}^\dagger\hat b^\dagger,\; \hat K_- = \hat a_{\mathrm{R}}\,\hat b,\, \mathrm{and} \, \hat K_0 = \tfrac{1}{2}(\hat a_{\mathrm{R}}^\dagger\hat a_{\mathrm{R}} + \hat b^\dagger\hat b + 1)$, giving $\hat{\mathcal{H}}_\mathrm{CRI}
		= \hbar\tilde\Omega\,\hat N_+ + \hbar\tilde\Delta\,\hat N_-
		- 2\hbar\bar g\,\hat K_1$ where $\hat N_\pm = \hat a_{\mathrm{R}}^\dagger\hat a_{\mathrm{R}} \pm \hat b^\dagger\hat b$, the Hermitian generator $\hat K_1 = \tfrac{1}{2}(\hat K_+ + \hat K_-)$, $\tilde\Omega = \tfrac{1}{2}(\Omega + \omega_{\rm cyc})$, and $\tilde\Delta = \tfrac{1}{2}(\Omega - \omega_{\rm cyc})$. Here, $\hat n_a = \hat a_{\mathrm R}^\dagger\hat a_{\mathrm R}$ and $\hat n_b = \hat b^\dagger\hat b$ are, respectively, the photon and matter number operators, with respective eigenvalues $n_a, n_b \in \mathbb{Z}_{\ge0}$. Since $[\hat{ \mathcal{H}}_{\rm CRI},\hat N_-]=0$, the difference $q = n_a - n_b \in \mathbb{Z}$ is conserved and labels invariant SU(1,1) symmetry sectors. Then the chiral-vacuum-dressed modes carrying different $q$ quantum numbers exhibit symmetry-protected level crossings (Fig.~\ref{Fig: fig1}b). The interaction creates and destroys photon--matter pairs, so $\hat N_+$ changes by $\pm 2$. The $q=0$ ground state is therefore an infinite superposition, $\ket{0}_{\rm CRI} = \sum_{p=0}^\infty C_p \ket{p, p}$, carrying finite photon and matter populations ($\langle \hat a_{\mathrm{R}}^\dagger\hat a_{\mathrm{R}} \rangle, \langle \hat b^\dagger\hat b \rangle > 0$) and squeezing correlations ($\langle \hat a_{\mathrm{R}}\hat b \rangle \neq 0$).
		
		The co-rotating terms are instead expressed in terms of the SU(2) algebra generators $\hat L_+ = \hat a_{\mathrm{L}}^\dagger\hat b,\;\hat L_- = \hat a_{\mathrm{L}}\,\hat b^\dagger,\, \mathrm{and} \,\hat L_0 = \tfrac{1}{2}(\hat a_{\mathrm{L}}^\dagger\hat a_{\mathrm{L}} - \hat b^\dagger\hat b)$, giving $\hat{\mathcal{H}}_\mathrm{CRA}
		= \hbar\tilde\Omega\,\hat N_+ + \hbar\tilde\Delta\,\hat N_-
		- 2\hbar\bar g\,\hat L_1$, where $\hat L_1 = (\hat L_+ + \hat L_-)/2$. Here, $[\hat{\mathcal{H}}_{\rm CRA},\hat N_+]=0$, the excitation number is conserved, each $Q = n_a + n_b \in \mathbb{Z}_{\geq 0}$ sector is finite $(Q+1)$-dimensional, and the $Q=0$ ground state is the bare vacuum $\ket{0_a, 0_b}$. 
		
		\textit{2DEG transport in chiral cavity---}The gauge-invariant current in the CM frame is $\hat{\boldsymbol{J}} = -(e\sqrt{N}/m^*)(\hat{\boldsymbol{\Pi}} + e\sqrt{N}\,\hat{\boldsymbol{A}}_{\textrm{cav}})$. 
		In terms of the Landau ladder and cavity photon operators, the Cartesian components of the current are expressed as
		\begin{align}
			\hat J_x & = J_0 [(\hat b+\hat b^\dagger ) - \zeta (\hat a_{\mathrm{L/R}}+\hat a_{\mathrm{L/R}}^\dagger )],\\
			\hat J_y &= -i J_0 [(\hat b-\hat b^\dagger ) + s\,\zeta (\hat a_{\mathrm{L/R}}-\hat a_{\mathrm{L/R}}^\dagger ) ],
			\label{Eq: J in a and b}
		\end{align}
		where $J_0 = \sqrt{N\,e^2\hbar\omega_{\mathrm{cyc}}/2m^*}$, $\zeta = \bar g/\omega_{\rm cyc}$, $s = -1$ (for CRA), and $s = +1$ (for CRI). 
		The DC conductivity tensor is obtained from the Kubo formula~\cite{kuboStatisticalMechanicalTheoryIrreversible1957}, 
		$\sigma_{a b}(\omega)=i (\omega+i \gamma)^{-1} \left[(e^2 n_{\mathrm{2D}}/m^*)\, \delta_{a b}+ S^{-1} \chi_{a b}(\omega) \right],$ with $a, b\in\{x, y\}$, 
		where the retarded current response $\chi_{ab}(\omega) = -\frac{i}{\hbar}\int_{-\infty}^{\infty} dt\; e^{i(\omega+i\gamma)t}\,\Theta(t)\,\langle[\hat J_a(t), \hat J_b(0)]\rangle$, $S$ is the surface area of the 2DEG, and $\gamma$ is a phenomenological broadening parameter for the coupled system~\cite{rokajWeakened, BartoloCiuti2018}. 
		
		Following Ref.~\cite{BartoloCiuti2018}, the Shubnikov--de Haas (SdH) oscillations enter through the field-dependent broadening parameter $\gamma_r(B)= W_{\mathrm{e}, r}\,\tau_\mathrm{e}^{-1}(B) + W_{\mathrm{p}, r}\,\tau_\mathrm{p}^{-1}$, where $W_{\mathrm{e}, r},\, W_{\mathrm{p}, r}$ are the electronic and photonic weights of mode $r$ (SM), and $\tau_\mathrm{p}$ is the transport scattering time of the photonic component, set by environmental fluctuations of the cavity mode rather than the photon lifetime. From a standard model for the density of states~\cite{AndoFowlerStern1982, Ihn2009}, we have the electronic scattering rate $\tau_\mathrm{e}^{-1}(B) = \tau_0^{-1} \left[1 - 2\exp\left(-\pi/(\omega_{\rm cyc} \tau_\mathrm{q}) \right)\cos(2\pi \nu)\right]$, where $\tau_0$ is the zero-field transport time, and $\tau_\mathrm{q}$ is the quantum lifetime, describing the scattering induced broadening of electron momentum states. For simulations, we assume $\omega_\mathrm{cav}\tau_0  = 100 $, $\omega_\mathrm{cav} \tau_\mathrm{q} = 2$, and $\omega_\mathrm{cav} \tau_\mathrm{p} = 1000$.  
		
		The cavity modifications to transport properties become more apparent in the resistivity tensor. By inverting the conductivity tensor, at constant broadening $\gamma$ and in the limit $\omega_\mathrm{cyc}\gg \gamma$, we obtain the approximate forms
		\begin{align}
			\frac{\rho_{xx}}{\rho_{\rm D}} \approx 1 + n_\mathrm{c} \frac{g^2}{\omega_{\rm cav}^2},\quad &
			\frac{\rho_{xy}^{(s)}}{h/\nu e^2} \approx 1 - s\,\frac{\gamma^2 g^2}{\omega_{\rm cav}^3\,\omega_{\rm cyc}},
			\label{Eq: rho chiral} 
		\end{align}
		where $n_\mathrm{c} =1$ for a chiral cavity, $n_\mathrm{c} = 2$ for an achiral cavity, and $\rho_{\rm D} = m^*\gamma/(n_{\rm 2D}e^2)$ is the classical Drude resistivity at $B=0$. The full details for the transport computations are provided in the SM.

		The SdH oscillations are modeled through the field-dependent broadening parameter $\gamma_r(B)$. Figs.\,\ref{Fig: fig1}(c) and (g) reveal a sharp contrast in longitudinal resistivity between the two helicities. In both cases, $\rho_{xx}/\rho_\mathrm{D}$ is offset by $g^2/\omega_\mathrm{cav}^2$ at $B=0$, where $\rho_\mathrm{D}$ is evaluated at $\gamma(0)=1/\tau_\mathrm{e}$. Because $\gamma_r(B)$ depends on electronic and photonic weights, light--matter hybridization modifies the SdH oscillations. The amplitude of the oscillations is strongly damped for the CRA case at zero detuning, owing to hybridization between the cavity mode and the collective cyclotron resonance, reducing the electronic weight. However, for the CRI case, the absence of coherent energy exchange in antiresonant interaction results in largely undamped SdH oscillations.
		
		The Hall resistivity is robust to dissipation and depends only on the external field. Dissipation is time-reversal symmetry-even and therefore enters only in the longitudinal transport channel. This picture remains valid for a 2DEG embedded in an achiral cavity where the current in Eq.~\eqref{Eq: J in a and b} contains cavity contributions from both polarizations, weighted by $\zeta = \bar g/\omega_{\rm cyc}$. The two circular polarizations contribute with opposite handedness, so their Hall contributions cancel, leading to a robust quantized Hall resistivity at zero temperature~\cite{RokajResistivity2026}.
		In a chiral cavity, however, the circularly polarized photons contribute to the current with a fixed handedness, adding to or subtracting from the cyclotron motion. This contribution, no longer compensated by the opposite polarization, appears in the Hall resistivity as a handedness ($s$)-dependent correction; see Eq.~\eqref{Eq: rho chiral}. 
		
		The chiral cavity-induced correction grows with increased light--matter coupling strength ($g$) and broadening parameter $(\gamma)$, as shown in Figs.~\ref{Fig: fig1}(d, h). The resulting deviation from the quantized Hall resistivity appears only for the chiral cavity. Importantly, the correction vanishes for $\gamma=0$, independent of $g$, indicating that a finite hybrid-mode lifetime is also required for the cavity to modify the dc Hall response. In our model, the chiral field remains present in the $B\to 0$ limit, and the response corresponds to free electrons coupled to chiral cavity modes~\cite{vacuumanomalousHall}. In this limit, we obtain a finite Hall resistivity $\rho^{(s)}_{xy}=-s\rho_D \gamma g^2/\omega^3_{\rm cav}$, 
		which is consistent with the anomalous Hall effect due to the circularly polarized photons as also found in Ref.~\cite{vacuumanomalousHall}. This zero-field response can be viewed as a vacuum analog of the light-induced anomalous Hall effect~\cite{McIveranomalousgraphene}. The anomalous Hall resistivity is found to vanish for zero broadening. 
		
		\textit{Current fluctuations and ground state squeezing---}In a Hall measurement (current sourced along $x$), the transverse current $\hat J_y$ is probed, which is a linear superposition of matter and photon quadrature operators in Eq.~\eqref{Eq: J in a and b}. Therefore, fluctuations in $\hat J_y$ probe the ground-state correlations of the coupled system~\cite{Passetti2023PRL}. Through $\hat c \equiv (\hat b + s\zeta\hat a_{\mathrm{L/R}})/\sqrt{1+\zeta^2}$, which satisfies $[\hat c,\hat c^\dagger]=1$, the normalized current quadrature $\hat j_y=(\hat c-\hat c^\dagger)/2i$ has variance $1/4$ for the bare vacuum. For the achiral case, $1+\zeta^2$ is replaced by $1+2\zeta^2$ to account for both cavity photon polarizations.
		
\begin{figure}[t!]
			\centering
			\includegraphics[width=0.9\linewidth]{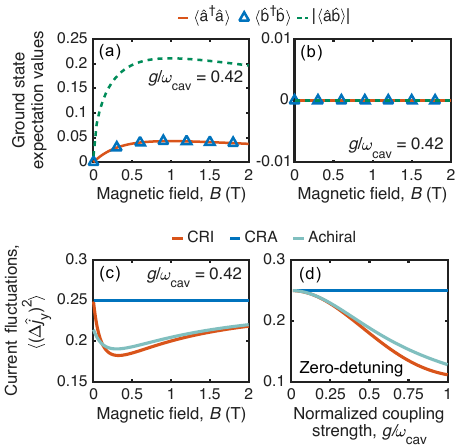}
			\caption{(a), (b)~ Photon $\langle a^{\dagger}a \rangle$ and matter $\langle b^{\dagger} b\rangle$ ground state populations and squeezing correlation $\langle\hat a\hat b\rangle$ of the CRI and CRA systems, respectively, finite for the two-mode squeezed vacuum. CRA ground state populations vanish identically, consistent with the bare vacuum.
			(c)~Current fluctuations versus magnetic field for $g/\omega_{\rm cav}=0.42$. The CRI current fluctuations fall below the bare vacuum limit ($1/4$), reflecting the two-mode squeezed ground state, while the CRA case remains at $1/4$. 
			(d)~Current fluctuations at zero-detuning versus coupling, $g/\omega_{\rm cav}$.}
			\label{Fig: fig2}
		\end{figure}

		Figs.~\ref{Fig: fig2}(c, d) show $\langle(\Delta j_y)^2\rangle$ as a function of magnetic field. The CRI current fluctuations fall below $1/4$, signifying two-mode squeezing of the SU(1,1) dressed ground state. This suppression originates from the virtual light--matter excitations generated by the counter-rotating interaction. In contrast, the CRA current variance remains at $1/4$, consistent with the SU(2) symmetry preserving the bare vacuum ground state. Figs.~\ref{Fig: fig2}(a, b) describe the ground state photon and matter populations and squeezing correlations of the CRI and CRA systems, respectively, linking their symmetries to the ground state properties.
		The achiral cavity also exhibits subvacuum fluctuations, arising from two sources: the counter-rotating interaction and the cross-polarization coupling in the diamagnetic terms, $\hbar D(a_{\rm L}a_{\rm R}+a_{\rm L}^\dagger a_{\rm R}^\dagger)$. Chiral cavities remove the latter processes, isolating the subvacuum fluctuations associated with the counter-rotating interaction~\cite{MakiharaEtAl2021NC,ciutiQuantumVacuumProperties2005}.
		
		The ground state current fluctuations can be inferred indirectly through the fluctuation--dissipation theorem (FDT), which relates the current-noise spectrum at thermal equilibrium to the dissipative current response encoded in the optical conductivity of the 2DEG--cavity system~\cite{ClerkEtAl2010RMP, Hauke2016NP}. Phase- and polarization-resolved THz spectroscopy at cryogenic temperatures can be used to determine the complex transmission spectrum. Using an appropriate input--output description, the relevant optical conductivity can be extracted and, through the FDT, used to reconstruct $\langle(\Delta j_y)^2\rangle$. Importantly, this scheme does not rely on directly measuring the current fluctuations through noise spectroscopy. Instead, the ground state correlations are inferred from the equilibrium linear response. 
		
\begin{figure}[t]
			\centering
			\includegraphics[width=0.8\linewidth]{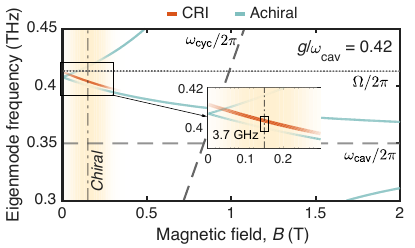}
			\caption{Predicted THz eigenspectrum for the chiral cavity of Ref.~\cite{kulkarniRealizationChiralPhotoniccrystal2026}. The chiral cavity, designed for $B=0.15$\,T remains functional over $B=0.05$--$0.25$\,T (shaded region). Within this range, it follows the CRI Hamiltonian and exhibits a single chiral-vacuum-dressed cavity mode shifted by $\sim$3.7\, GHz; outside it, the achiral three-mode spectrum is recovered. The inset zooms in on the CRI mode at the operational field. The parameters are $\omega_{\rm cav}/2\pi=0.35$\,THz with a normalized coupling strength of $g/\omega_\mathrm{cav}\sim 0.42$, and a 10-quantum-well GaAs heterostructure with an electron density of $5\times10^{11}~\mathrm{cm}^{-2}$ per quantum well, and an effective electron mass $m^\ast=0.067\,m_\mathrm{e}$ (SM). 			The dashed, dotted, and dash-dotted lines mark the cavity frequency, the renormalized cavity frequency $\Omega$, and the cyclotron frequency, respectively.}
			\label{Fig: fig3}
		\end{figure}

		\textit{Spectroscopic prediction for a chiral cavity---}A photonic crystal cavity built from gyrotropic layers has recently been implemented to confine light of one handedness while absorbing the other at THz frequencies~\cite{kulkarniRealizationChiralPhotoniccrystal2026}, following the design of Ref.~\cite{tayTerahertzChiralPhotoniccrystal2025a}. We present the eigenmodes of a 2DEG embedded in such a cavity, Fig.~\ref{Fig: fig3}, which can be measured with THz time-domain spectroscopy.  
		
		In an achiral cavity, the vector potential contains both circular polarizations, so the diamagnetic term couples the two helicities, $\hat{\boldsymbol{A}}^2_\mathrm{cav} \propto (\hat a_{\mathrm{L}} + \hat a^\dagger_{\mathrm{R}}) (\hat a_{\mathrm{R}} + \hat a^\dagger_{\mathrm{L}})$. In a chiral cavity, $\boldsymbol{e}_{\mathrm{L/R}} \cdot \boldsymbol{e}_{\mathrm{L/R}} = 0$ simplifies $\hat{\boldsymbol{A}}^2_\mathrm{cav}$  to $\hat a_{\mathrm{R}}^\dagger \hat a_{\mathrm{R}}$. Consequently, the renormalized cavity frequency for $B \to  0$ stays at $\sqrt{\omega_{\rm cav}^2 + 2g^2}$ for the achiral mode and at $\Omega$ for the chiral mode~\cite{liVacuumBlochSiegert2018, yang2025quantum}.
		
		In a CRI-only cavity, only the chiral-vacuum-dressed cavity mode is bright (highlighted in red in Fig.~\ref{Fig: fig3}, see SM).  
	 	In a CRA-only cavity, both branches are bright. The chiral cavity, designed for producing a chiral mode at $B=0.15$~T, maintains chiral operation over $0.05 < B < 0.25$~T, confining only the right-circularly polarized mode, and reverts to achiral operation beyond this range~\cite{tayTerahertzChiralPhotoniccrystal2025a, kulkarniRealizationChiralPhotoniccrystal2026}. Within this window, we model the cavity with a single right-circularly polarized mode, where the light--matter interaction follows the CRI Hamiltonian: the spectrum collapses from three achiral modes~\cite{liVacuumBlochSiegert2018} to the single bright chiral-vacuum-dressed cavity mode, offset by $\sim 3.7$~GHz from the achiral cavity-like branch, shown in Fig.~\ref{Fig: fig3}.
		
		\textit{Discussion and outlook---}The chiral cavity separates the CRA and CRI contributions, allowing their spectroscopic and transport signatures to be distinguished experimentally. The softening of the lower hybrid modes, particularly in CRA-cavity, reduces the thermal activation gap and leads to finite-temperature corrections~\cite{rokajWeakened}. Beyond the Kohn limit, disorder-assisted cavity hopping provides an additional scattering channel through antiresonant interactions, which is present in CRI-cavity but absent in CRA-cavity. Therefore, chiral cavities provide a platform to directly test the modified Hall transport predicted to be induced by cavity-mediated disorder hopping~\cite{ciutiCavitymediatedElectronHopping2021,arwasQuantumElectronTransport2023}. 
	    
		The squeezed CRI ground state can be probed through current fluctuations, providing a physical observable of nonclassical correlations beyond the rotating-wave approximation. 
		These fluctuations can be measured directly through current-noise spectroscopy or inferred indirectly from the optical response of the system via the fluctuation-dissipation theorem. Although the linear spectrum of the coupled system can be reproduced by both classical and quantum descriptions~\cite{Hughes2024OQ}, the optical response retains the underlying quantum correlations. The SU(1,1) structure motivates further studies of virtual photon emission under nonadiabatic modulation ~\cite{deLiberatoQuantumVacuumRadiation2007,hagenmullerAllopticalDynamicalCasimir2016,gunterSubcycleSwitchonUltrastrong2009} in the CRI-cavity. Extending the model to spatially varying cavity fields and disorder would enable studies on the breakdown of Kohn's theorem and cavity-induced modifications of the quantum Hall phases~\cite{EnknerEtAl2025N,baccioniTheoryFractionalQuantum2025, dag2026quantumhallferromagnetismcavity} in an electromagnetic environment where the handedness is a control parameter for light--matter interactions.

		\textit{Acknowledgments---}The authors thank Kaden R. Hazzard for fruitful discussions. V.R. acknowledges the financial support provided by the Villanova University Summer Grant program during the summer of 2026. J.K.\ acknowledges support from the U.S.\ Army Research Office (through Award No.\ W911NF-25-2-0150, W911NF-21-1-0157), the Gordon and Betty Moore Foundation (through Grant No.\ 11520), and the Robert A.\ Welch Foundation (through Grant No.\ C-1509). MB is supported by the Japan Society for the Promotion of Science (JSPS) (Grant Nos. JP24K21526, JP25K00012, JP25K01691, JP25K01694, JP26K01332, JPJSJRP20221202) and the Research Foundation for Opto-Science and Technology. This research was supported in part by grant NSF PHY2309135 to the Kavli Institute for Theoretical Physics (KITP).
	
		\bibliography{references}

	\end{document}


\setcounter{figure}{0}
\renewcommand{\thefigure}{S\arabic{figure}}

\setcounter{equation}{0}
\renewcommand{\theequation}{S\arabic{equation}}
\newpage
\clearpage
\onecolumngrid

	\begin{center}
		{\Large\bfseries Supplementary Material}
	\end{center}
	
	Our model considers a two-dimensional electron gas coupled to a strong magnetic field and a circularly polarized homogeneous cavity field. The system is described by the minimal-coupling Hamiltonian
	\begin{equation}\label{Eq: pauli-fierz}
		\hat{\mathcal{H}} = \sum^{N}_{i=1}\frac{\big(\hat{\bm{\pi}}_i+e\hat{\boldsymbol{A}}\big)^2}{2m^*} + \hbar\omega_{\mathrm{cav}}\left(\hat{a}_{\mathrm{L/R}}^{\dagger}\hat{a}_{\mathrm{L/R}} + \frac{1}{2}\right) + \sum_{i<j}W(\boldsymbol{r}_i-\boldsymbol{r}_j),
	\end{equation}
	where $\hat{\bm{\pi}}_i=i\hbar\nabla_i+e\boldsymbol{A}_{\mathrm{ext}}(\boldsymbol{r}_i)$ is the kinetic momenta of the electrons, $\boldsymbol{A}_{\mathrm{ext}}(\boldsymbol{r})=-\hat{\boldsymbol{x}}\,By$ describes the applied magnetic field $\boldsymbol{B}=\nabla \times \boldsymbol{A}_{\mathrm{ext}}(\boldsymbol{r})=B\hat{\boldsymbol{z}}$, effective mass $m^*$, and cavity frequency $\omega_{\mathrm{cav}}$. With Galilean invariance in a homogeneous system, the center-of-mass (CM) is decoupled from the relative motion of the electrons.
	
	The cavity field is circularly polarized:
	\begin{equation}\label{Eq: cavity field}
		\hat{\boldsymbol{A}} = A_0\left(\boldsymbol{e}_{\mathrm{L/R}}\,\hat{a}_{\mathrm{L/R}} + \boldsymbol{e}_{\mathrm{L/R}}^*\,\hat{a}_{\mathrm{L/R}}^{\dagger}\right), \quad A_0 = \sqrt{\frac{\hbar}{2\epsilon_r\epsilon_0\mathcal{V}\omega_{\mathrm{cav}}}},
	\end{equation}
	where $\boldsymbol{e}$ is the polarization vector, $\mathcal{V}$ is the effective mode volume, $\epsilon_0$ is the vacuum permittivity, and $\epsilon_r$ is the relative permittivity of the background medium. The operators $\hat{a}$ and $\hat{a}^{\dagger}$ satisfy $[\hat{a},\hat{a}^{\dagger}]=1$, and $W(\boldsymbol{r}_i-\boldsymbol{r}_j)$ is the Coulomb interaction. The circular polarization vectors are defined as:
	\begin{equation*}
		\text{Right-circular (RCP):} \quad \boldsymbol{e}_{\mathrm{R}} = \frac{1}{\sqrt{2}}(1,\,i), \qquad
		\text{Left-circular (LCP):} \quad \boldsymbol{e}_{\mathrm{L}} = \frac{1}{\sqrt{2}}(1,\,-i). 
	\end{equation*}

	\subsection{Center-of-mass Hamiltonian}
	\label{sec:cm hamiltonian}
	
	Further, $W(\boldsymbol{r}_i-\boldsymbol{r}_j)=1/4\pi \epsilon_0 |\boldsymbol{r}_i-\boldsymbol{r}_j|$ is the Coulomb interaction between the electrons. We have parameterized the bare electron dispersion by an effective mass $m$ and assumed Galilean invariance. With Galilean invariance in a homogeneous system, the CM is decoupled from the relative motion of the electrons, regardless of the interaction strength~\cite{Kohn1961}. The kinetics of the CM and its coupling to light is best described in terms of the CM coordinate $\boldsymbol{R}=(X,Y)=\sum^N_{i=1}\boldsymbol{r}_i/\sqrt{N}$ where $N$ is the total particle number. Following the derivation presented in the Supplementary Material (SM), we obtain the Hamiltonian describing the coupling of the CM to light \\
	\begin{equation}
		\hat{\mathcal{H}}_{\textrm{CM}}=\frac{1}{2m^*}\left(\hat{\bm{\Pi}}+e\sqrt{N}\hat{\boldsymbol{A}}\right)^2+\hbar\omega_{\mathrm{cav}}\left(\hat{a}^{\dagger}_\mathrm{R}\hat{a}_\mathrm{R}+\frac{1}{2}\right)    
	\end{equation}
	
	where $\hat{\bm{\Pi}}=\textrm{i}\hbar \nabla_{\boldsymbol{R}}+e\boldsymbol{A}_{\textrm{ext}}(\boldsymbol{R})$ is the dynamical momentum of the CM. After expanding the CM Hamiltonian, we have
	\begin{equation}
		\hat{\mathcal{H}}_{\textrm{CM}}=\frac{\hat{\bm{\Pi}}^2}{2m^*}+\frac{e\sqrt{N}}{m^*}\hat{\bm{\Pi}} \cdot \hat{\boldsymbol{A}} + \frac{e^2N}{2m^*}\hat{\boldsymbol{A}}^2+  \hbar\omega_{\mathrm{cav}}\left(\hat{a}^{\dagger}_\mathrm{R}\hat{a}_\mathrm{R}+\frac{1}{2}\right).
	\end{equation}
	Substituting the expression for the chiral cavity field $\hat{\boldsymbol{A}}$, we can bring the diamagnetic contribution $\hat{\boldsymbol{A}}^2$ into a diagonal form of the chiral photon mode
	\begin{equation}
		\hat{\mathcal{H}}_{\textrm{CM}}=\frac{\hat{\bm{\Pi}}^2}{2m^*}+\frac{e\sqrt{N}}{m^*}\hat{\bm{\Pi}} \cdot \hat{\boldsymbol{A}} + \hbar D\,\hat{a}^{\dagger}_\mathrm{R}\hat{a}_\mathrm{R}+  \hbar\omega \hat{a}^{\dagger}_\mathrm{R}\hat{a}_\mathrm{R}.
	\end{equation}
	Note that above we dropped the trivial constant terms and we introduced the diamagnetic frequency $D=e^2N/(2m^*\epsilon_0\epsilon_r\mathcal{V} \omega_{\mathrm{cav}}) = g^2/\omega_{\mathrm{cav}}$. Here, $g$ is the collective light--matter coupling strength given by, $g = \sqrt{\frac{Ne^2}{2\epsilon_r\epsilon_0\mathcal{V}m^*}}$. Due to the Landau gauge in the $x$ direction, the Hamiltonian is translationally invariant, thus the eigenstates are plane-waves of the form $e^{\mathrm{i}KX}$. We apply the Hamiltonian on the plane wave, and after introducing the shifted electronic coordinates $\bar Y=Y+\hbar K/eB$, the Hamiltonian takes the form
	\begin{equation}
		\hat{\mathcal{H}}_{\textrm{CM}}=-\frac{\hbar^2}{2m^*} \partial^2_{\bar Y}+\frac{m^*\omega^2_\mathrm{cyc}}{2}\bar Y^2-\frac{e^2\sqrt{N}A_0B}{m^*\sqrt{2}}\bar Y (\hat{a}_\mathrm{R}+\hat{a}^{\dagger}_\mathrm{R})-\frac{\hbar e \sqrt{N}A_0}{m^*\sqrt{2}} (\hat{a}_\mathrm{R}-\hat{a}^{\dagger}_\mathrm{R})\partial_{\bar Y}+\hbar\Omega \hat{a}^{\dagger}_\mathrm{R}\hat{a}_\mathrm{R},\;\; \textrm{where}\;\; \Omega=\omega_{\mathrm{cav}} + D.
	\end{equation}
	
	Finally, we write the Landau level part of the Hamiltonian in terms of annihilation and creation operators,
	\begin{equation}\label{Eq: LL operators}
		\bar Y= \sqrt{\frac{\hbar}{2m^*\omega_{\mathrm{cyc}}}} \left(\hat{b} +\hat{b}^{\dagger}\right) \;\; \textrm{and}\;\; \partial_{\bar Y}=\sqrt{\frac{m^* \omega_{\mathrm{cyc}}}{2\hbar}}\left(\hat{b}-\hat{b}^{\dagger}\right)
	\end{equation}
	where the operators satisfy bosonic commutation relations $[\hat{b},\hat{b}^{\dagger}]=1$. In terms of the bosonic operators, the light-matter Hamiltonian simplifies significantly
	\begin{equation}
		\hat{\mathcal{H}}_{\textrm{CRI}}=\hbar \omega_{\mathrm{cyc}} \hat{b}^{\dagger}\hat{b} + \hbar \Omega \hat{a}^{\dagger}_\mathrm{R} \hat{a}_\mathrm{R} -\hbar \bar g\left(\hat{b}\hat{a}_\mathrm{R} + \hat{b}^{\dagger}\hat{a}^{\dagger}_\mathrm{R}\right),\; \mathrm{where} \;\bar g = g\sqrt{\omega_{\mathrm{cyc}}/\omega_{\mathrm{cav}}}
	\end{equation}
	
	Diagonalizing the above Hamiltonian using the Bogoliubov transformation, the polariton annihilation operator that diagonalizes the system is expressed as:
	
	\begin{equation}
		\label{Eq: Chiral pr operator}
		\hat{p}_r = w_r \hat{a}_{R} + x_r \hat{b} + y_r \hat{a}_\mathrm{R}^\dagger + z_r \hat{b}^\dagger
	\end{equation}
	
	Representing the Hopfield coefficients as \(\vec{v}_r = (w_r, x_r, y_r, z_r)^T\), the Hamiltonian reduces to a simple eigenvalue problem.
	
	\begin{equation}\label{Eq: m cri}
		M_{\mathrm{CRI}} = \begin{pmatrix}
			\Omega & 0 & 0 & \bar{g} \\
			0 & \omega_{\mathrm{cyc}} & \bar{g} & 0 \\
			0 & -\bar{g} & -\Omega & 0 \\
			-\bar{g} & 0 & 0 & -\omega_{\mathrm{cyc}}
		\end{pmatrix}.
	\end{equation}
	This matrix has a block structure coupling $(w,z)$ and $(x,y)$ independently. The positive eigenvalues are
	\begin{equation}\label{Eq: omega cri}
		\Omega_\pm^{\mathrm{CRI}} = \frac{\Sigma \pm \Delta}{2},
	\end{equation}
	where $\Delta = \Omega - \omega_{\mathrm{cyc}}$ is the detuning and $\Sigma = \sqrt{(\Omega + \omega_{\mathrm{cyc}})^2 - 4\bar{g}^2}$. 
	
	The Hopfield coefficients, obtained from the eigenvectors of Eq.~\ref{Eq: m cri} are
	\begin{align}
		\left(\begin{array}{c}
			w^{\mathrm{CRI}}_+\\
			x^{\mathrm{CRI}}_+\\
			y^{\mathrm{CRI}}_+\\
			z^{\mathrm{CRI}}_+\\
		\end{array}\right) = \left(\begin{array}{c}
			\cosh\, r\\
			0\\
			0\\
			-\sinh\, r\\
		\end{array}\right) \quad \mathrm{ and }\quad 
		\left(\begin{array}{c}
			w^{\mathrm{CRI}}_-\\
			x^{\mathrm{CRI}}_-\\
			y^{\mathrm{CRI}}_-\\
			z^{\mathrm{CRI}}_-\\
		\end{array}\right) = \left(\begin{array}{c}
			0\\
			\cosh\, r\\
			-\sinh\, r\\
			0\\
		\end{array}\right),
	\end{align}
	where $\tanh\, 2r = \frac{2\bar g}{\Omega + \omega_{\mathrm{cyc}}}$.
	
	The photonic and electronic weight of mode $r$ is given by:
	\begin{equation}
		\label{Eq: Wer Wpr}
		W_{\rm e, r} = |x_r|^2 - |z_r|^2, \qquad
		W_{\rm p, r} = |w_r|^2 - |y_r|^2.
	\end{equation}

	\subsection{Cavity confining left-circular photon field}
	Along the same lines as above, we represent the left-circular cavity field with
	\begin{equation}\label{Eq: a cartesian lcp}
		\boldsymbol{e}_L = (1,-i)/\sqrt{2}; \qquad \hat{A}_x = \frac{A_0}{\sqrt{2}}\left(\hat{a}_L + \hat{a}_L^\dagger\right), \qquad \hat{A}_y = \frac{-iA_0}{\sqrt{2}}\left(\hat{a}_L - \hat{a}_L^\dagger\right).
	\end{equation}
	
	The center-of-mass Hamiltonian with the shifted electronic coordinates can then be written as
	\begin{equation}\label{Eq: hcm ybar lcp}
		\begin{aligned}
			\hat{\mathcal{H}}_{\textrm{CM}} &= -\frac{\hbar^2}{2m^*}\partial_{\bar{Y}}^2 + \frac{m^*\omega_{\mathrm{cyc}}^2}{2}\bar{Y}^2 - \frac{e^2\sqrt{N}\,B\,A_0}{m^*\sqrt{2}}\,\bar{Y}\left(\hat{a}_L + \hat{a}_L^\dagger\right) + \frac{\hbar e\sqrt{N}\,A_0}{m^*\sqrt{2}}\,\partial_{\bar{Y}}\left(\hat{a}_L - \hat{a}_L^\dagger\right) + \hbar\Omega\,\hat{a}_L^\dagger \hat{a}_L.
		\end{aligned}
	\end{equation}
	Thus, using the Landau level operators defined in \ref{Eq: LL operators}, we can reduce the CM Hamiltonian to 
	\begin{equation}\label{Eq: h cra}
		\hat{\mathcal{H}}_{\textrm{CRA}} = \hbar\omega_{\mathrm{cyc}}\,\hat{b}^\dagger \hat{b} + \hbar\Omega\,\hat{a}_L^\dagger \hat{a}_L - \hbar\bar{g}\left(\hat{b}^\dagger \hat{a}_L + \hat{b}\,\hat{a}_L^\dagger\right).
	\end{equation}
	
	The Hopfield matrix for the CRA Hamiltonian is
	\begin{equation}\label{Eq: m cra}
		M_{\mathrm{CRA}} = \begin{pmatrix}
			\Omega & -\bar{g} & 0 & 0 \\
			-\bar{g} & \omega_{\mathrm{cyc}} & 0 & 0 \\
			0 & 0 & -\Omega & \bar{g} \\
			0 & 0 & \bar{g} & -\omega_{\mathrm{cyc}}
		\end{pmatrix}.
	\end{equation}
	Block-diagonal: the upper $2\times 2$ gives the positive eigenvalues
	\begin{equation}\label{Eq: omega cra}
		\Omega_\pm^{\mathrm{CRA}} = \frac{(\Omega + \omega_{\mathrm{cyc}}) \pm \sqrt{(\Omega - \omega_{\mathrm{cyc}})^2 + 4\bar{g}^2}}{2}.
	\end{equation}
	For CRA, $y_r = z_r = 0$ for all branches. The eigenvectors are given by
	\begin{align}
		\left(\begin{array}{c}
			w^{\mathrm{CRA}}_+\\
			x^{\mathrm{CRA}}_+\\
			y^{\mathrm{CRA}}_+\\
			z^{\mathrm{CRA}}_+\\
		\end{array}\right) = \left(\begin{array}{c}
			\cos\, \theta\\
			-\sin\, \theta\\
			0\\
			0\\
		\end{array}\right) \quad \mathrm{ and }\quad 
		\left(\begin{array}{c}
			w^{\mathrm{CRA}}_-\\
			x^{\mathrm{CRA}}_-\\
			y^{\mathrm{CRA}}_-\\
			z^{\mathrm{CRA}}_-\\
		\end{array}\right) = \left(\begin{array}{c}
			\sin\, \theta\\
			\cos\, \theta\\
			0\\
			0\\
		\end{array}\right),
	\end{align}
	where $\tan\, 2\theta = \frac{2\bar g}{\Delta}$.
	
	\subsection{Achiral cavities}
	We consider an achiral cavity supporting both circular polarizations at the same frequency $\omega_{\mathrm{cav}}$. The vector potential is
	\begin{equation}\label{Eq: a achiral}
		\hat{\boldsymbol{A}} = A_0\left(\boldsymbol{e}_\mathrm{R}\,\hat{a}_\mathrm{R} + \boldsymbol{e}_\mathrm{R}^*\,\hat{a}_\mathrm{R}^\dagger + \boldsymbol{e}_L\,\hat{a}_\mathrm{L} + \boldsymbol{e}_L^*\,\hat{a}_\mathrm{L}^\dagger\right).
	\end{equation}
	
	Following the same calculation as above, we obtain the light--matter coupling Hamiltonian for a 2DEG confined in an achiral cavity
	\begin{equation}\label{Eq: h achiral bare}
		\mathcal{H}_{\mathrm{ach}} = \hbar\omega_{\mathrm{cav}}\left(\hat{a}_\mathrm{L}^\dagger \hat{a}_\mathrm{L} + \hat{a}_\mathrm{R}^\dagger \hat{a}_\mathrm{R}\right) + \hbar\omega_{\mathrm{cyc}}\,\hat{b}^\dagger \hat{b} - \hbar\bar{g}\left[\hat{b}^\dagger(\hat{a}_\mathrm{L} + \hat{a}_\mathrm{R}^\dagger) + \hat{b}(\hat{a}_\mathrm{R} + \hat{a}_\mathrm{L}^\dagger)\right] + \hbar D\left(\hat{a}_\mathrm{L} + \hat{a}_\mathrm{R}^\dagger\right)\left(\hat{a}_\mathrm{R} + \hat{a}_\mathrm{L}^\dagger\right).
	\end{equation}
	
	The polariton operators are $\hat{p}_r = \sum_{\xi = \pm} w_{r, \xi} \hat{a}_\xi + x_{r, \xi} \, \hat{b} + \sum_{\xi = \pm}y_{r, \xi} \hat{a}^\dagger_\xi + z_{r} \hat{b}^\dagger $. The polariton modes are obtained by diagonalizing the matrix:
	\begin{equation}\label{Eq: m linear}
		M_{\mathrm{achiral}} = \begin{pmatrix}
			\Omega & 0 & -\bar{g} & 0 & -D & 0 \\
			0 & \Omega & 0 & -D & 0 & \bar{g} \\
			-\bar{g} & 0 & \omega_{\mathrm{cyc}} & 0 & \bar{g} & 0 \\
			0 & D & 0 & -\Omega & 0 &\bar{g} \\
			D & 0 & -\bar{g} & 0 & -\Omega & 0 \\
			0 & -\bar{g} & 0 & \bar{g} & 0 & -\omega_{\mathrm{cyc}}
		\end{pmatrix}.
	\end{equation}
	
	\subsection{Inverse Bogoliubov transformation}
	\label{sec:inverse bogoliubov}
	
	For the CRA/CRI Hamiltonians, from the commutation relations, we have
	\begin{equation}
		\hat{a}_{\mathrm{L/R}} = \sum_r \bigl(w_r^*\,\hat{p}_r - y_r\,\hat{p}_r^\dagger\bigr), \quad \hat{b} = \sum_r \bigl(x_r^*\,\hat{p}_r - z_r\,\hat{p}_r^\dagger\bigr). \label{Eq: inverse ab}
	\end{equation}
	For the achiral case, 
	\begin{equation}
		\hat{a}_\xi = \sum_r \bigl(w_{r, \xi}^*\,\hat{p}_r - y_{r, \xi}\,\hat{p}_r^\dagger\bigr),\quad \hat{b} = \sum_r \bigl(x_r^*\,\hat{p}_r - z_r\,\hat{p}_r^\dagger\bigr). 
		\label{Eq: achiral inverse bogo}
	\end{equation}
	
	\section{Gauge Invariant Current and Conductivity}
	
	The velocity of each electron is obtained from the Heisenberg equation of motion:
	\begin{equation}\label{Eq: velocity}
		\hat{\boldsymbol{v}}_i = \frac{\mathrm{d}\boldsymbol{r}_i}{\mathrm{d}t} = \frac{i}{\hbar}[\mathcal{H}, \boldsymbol{r}_i] = \frac{1}{m^*}\left(-i\hbar\bm{\nabla}_i - e\boldsymbol{A}_{\mathrm{ext}}(\boldsymbol{r}_i) - e\hat{\boldsymbol{A}}\right) = \frac{1}{m^*}\left(\hat{\bm{\pi}}_i + e\hat{\boldsymbol{A}}\right).
	\end{equation}
	
	The total current is $\hat{\boldsymbol{J}} = -e\sum_{i=1}^N \hat{\boldsymbol{v}}_i$. Going to the CM frame using $\sum_{i=1}^N \bm{\nabla}_i = \sqrt{N}\bm{\nabla}_{\boldsymbol{R}}$ and $\sum_{i=1}^N \boldsymbol{A}_{\mathrm{ext}}(\boldsymbol{r}_i) = \sqrt{N}\boldsymbol{A}_{\mathrm{ext}}(\boldsymbol{R})$, the total current depends only on CM coordinates:
	\begin{equation}\label{Eq: total current}
		\hat{\boldsymbol{J}} = -\frac{e\sqrt{N}}{m^*}\left(\hat{\bm{\Pi}} + e\sqrt{N}\,\hat{\boldsymbol{A}}\right).
	\end{equation}
	This result, following from Kohn's theorem, implies that the relative coordinates of the electrons drop out of current response entirely~\cite{rokajWeakened}. Utlizing the Landau ladder operators \ref{Eq: LL operators}, we can represent the Cartesian components of current as 
	\begin{equation}
		\hat J_x = J_0\left[\left(\hat{b}+\hat{b}^\dagger\right) - \zeta\left(\hat{a}+\hat{a}^\dagger\right)\right], \quad \hat J_y = -\mathrm{i}J_0\left[\left(\hat{b}-\hat{b}^\dagger\right) + s\,\zeta\left(\hat{a}-\hat{a}^\dagger\right)\right], \label{Eq: jxjy}
	\end{equation}
	where
	\begin{equation*}
		J_0 = \sqrt{\frac{Ne^2\hbar\omega_{\mathrm{cyc}}}{2m^*}}, \qquad \zeta = \frac{\bar g}{\omega_{\mathrm{cyc}}}; \qquad s = \begin{cases} +1 & \text{CRI (RCP)}, \\ -1 & \text{CRA (LCP)}.\end{cases}
	\end{equation*}
	Here, $s$ captures the effect of the handedness of the cavity field interacting with the 2DEG. This asymmetry has direct consequences for the Hall conductivity. Substituting the inverse Bogoliubov transformation into Eqs.~\ref{Eq: jxjy}:
	\begin{equation}
		\label{Eq: J in p}
		\hat J_x = J_0\sum_r \left[\Lambda_{x,r}^*\,\hat{p}_r + \Lambda_{x,r}\,\hat{p}_r^\dagger\right], \quad
		\hat J_y = -\mathrm{i}J_0\sum_r \left[\Lambda_{y,r}^*\,\hat{p}_r - \Lambda_{y,r}\,\hat{p}_r^\dagger\right],
	\end{equation}
	where the current weights are
	\begin{equation}
		\begin{aligned}
			\Lambda_{x, r} = (x_r - z_r) - \zeta\,(w_r - y_r) ; \quad\Lambda_{y, r} = (x_r + z_r) + s\,\zeta\,(w_r + y_r) \\
		\end{aligned}
		\label{Eq: currentweight}
	\end{equation}
	
	Similarly, inserting Eqs~\ref{Eq: achiral inverse bogo} in Eqs.~\ref{Eq: J in p}, we obtain current weight $\Lambda^\mathrm{ach}_{x/y, r}$ for the achiral cavity case.
	
	The components of the current operator in terms of the polariton operators for RCP are
	\begin{eqnarray}
		\hat{J}^{\mathrm{CRI}}_x&=&J_0\left[\cosh r (\hat{p}_-+\hat{p}^{\dagger}_-) +\sinh r(\hat{p}_++\hat{p}_+^{\dagger})\right] -J_0\zeta \left[\cosh r (\hat{p}_+ +\hat{p}^{\dagger}_+) +\sinh r (\hat{p}_-+\hat{p}^{\dagger}_-)\right]\\
		\hat{J}^{\mathrm{CRI}}_y&=&-iJ_0\left[\cosh r(\hat{p}_--\hat{p}^{\dagger}_-)+\sinh r (\hat{p}^{\dagger}_+-\hat{p}_+)\right]-iJ_0\zeta\left[\cosh r(\hat{p}_+-\hat{p}^{\dagger}_+)+\sinh r (\hat{p}_--\hat{p}^{\dagger}_-)\right].
	\end{eqnarray}
	For the LCP case the components of the current are
	\begin{eqnarray}
		\hat{J}^{\mathrm{CRA}}_x&=&J_0\left[\cos\theta(\hat{p}_-+\hat{p}^{\dagger}_-)-\sin\theta (\hat{p}_++\hat{p}^{\dagger}_+)\right] -J_0\zeta\left[\cos\theta(\hat{p}_++\hat{p}^{\dagger}_+)+\sin\theta(\hat{p}_-+\hat{p}^{\dagger}_-)\right]\\
		\hat{J}^{\mathrm{CRA}}_y&=&-iJ_0\left[\cos\theta(\hat{p}_--\hat{p}^{\dagger}_-)+\sin\theta(\hat{p}^{\dagger}_+-\hat{p}_+)\right] +iJ_0\zeta\left[\cos\theta(\hat{p}_+-\hat{p}^{\dagger}_+)+\sin\theta(\hat{p}_--\hat{p}^{\dagger}_-)\right]
	\end{eqnarray}

	\section{Current-Current Response and the Kubo Formula}
	\label{sec:response}
	
	\subsection{Finite-temperature}
	\label{sec:finite temp}
	
	Following the discussion outlined in \cite{rokajWeakened}, the full eigenstates of $\mathcal{H} = \mathcal{H}_{\mathrm{cm}} + \mathcal{H}_{\mathrm{rel}}$ are product states $|\Psi_{n,I}\rangle = |\Phi_n\rangle \otimes |F_I\rangle$ with eigenvalues $E_{n,I} = E_n + E_I$, where $|\Phi_n\rangle$ are CM eigenstates and $|F_I\rangle$ are relative-coordinate eigenstates. Since $\hat{\boldsymbol{J}}$ depends only on CM operators (Eq.~\ref{Eq: total current}), the current matrix elements satisfy
	\begin{equation}\label{Eq: current separation}
		\langle \Psi_{n,I}|\hat{\boldsymbol{J}}|\Psi_{m,J}\rangle = \delta_{IJ}\langle \Phi_n|\hat{\boldsymbol{J}}|\Phi_m\rangle.
	\end{equation}
	The retarded current-current correlator is
	\begin{equation}\label{Eq: chi retarded}
		\chi_{ab}(t) = -\frac{i\,\Theta(t)}{\hbar}\langle [\hat J_a(t), \hat J_b]\rangle,
	\end{equation}
	where $\hat J_a(t) = e^{i\mathcal{H}t/\hbar}\hat J_a\,e^{-i\mathcal{H}t/\hbar}$. Using Eq.~\ref{Eq: current separation} and taking Fourier-transform, the correlator becomes
	\begin{equation}\label{Eq: chi fourier finite T}
		\chi_{ab}(\omega) = \frac{1}{Z_{\mathrm{CM}}}\sum_{n,m}\frac{e^{-\beta E_n} - e^{-\beta E_m}}{\omega + (E_n - E_m)/\hbar + i\gamma}\,\langle \Phi_n|\hat J_a|\Phi_m\rangle\langle \Phi_m|\hat J_b|\Phi_n\rangle,
	\end{equation}
	where $Z_{\mathrm{CM}} = \sum_n e^{-\beta E_n}$ is the CM partition function, $\beta = 1/(k_B T)$, and $\gamma > 0$ is a phenomenological broadening parameter.
	
	\subsection{Zero-temperature limit}
	\label{sec:zero temp}
	
	At $T = 0$, the ground state, $|\Phi_0\rangle$, is the polariton vacuum. Its contribution to current response becomes:
	\begin{equation}\label{Eq: chi T0}
		\chi_{ab}(\omega) = \sum_m \left[\frac{\langle 0|\hat J_a|m\rangle\langle m|\hat J_b|0\rangle}{\omega + (E_0 - E_m)/\hbar + i\gamma} - \frac{\langle 0|\hat J_b|m\rangle\langle m|\hat J_a|0\rangle}{\omega - (E_0 - E_m)/\hbar + i\gamma}\right].
	\end{equation}
	Since the current operators are linear in the polariton creation and annihilation operators (Eqs.~\ref{Eq: inverse ab}), only single-polariton excitations $|1_r\rangle = \hat{p}_r^\dagger|0\rangle$ contribute, with energies $\hbar\Omega_r$. The current response is 
	\begin{equation}
		\chi_{xx}(\omega) = \frac{J_0^2}{\hbar}\sum_r \frac{2\Omega_r\,|\Lambda_{x,r}|^2}{(\omega + i\gamma)^2 - \Omega_r^2}, \quad	\chi_{xy}(\omega) = \frac{iJ_0^2}{\hbar}\sum_r\left[\frac{\Lambda_{x,r}^*\,\Lambda_{y,r}}{\omega + i\gamma - \Omega_r} + \frac{\Lambda_{x,r}\,\Lambda_{y,r}^*}{\omega + i\gamma + \Omega_r}\right]. \label{Eq: chi}
	\end{equation}
	
	After substituting the expressions for the $\Lambda$'s we find for the CRI case
	\begin{eqnarray}
		\chi_{xx}(\omega)&=&\frac{2J^2_0}{\hbar}\left[\frac{\left(\cosh r -\zeta \sinh r\right)^2\Omega_-}{(\omega+i\gamma)^2-\Omega^2_-}+\frac{(\sinh r -\zeta \cosh r)^2 \Omega_+}{(\omega+\textrm{i}\gamma)^2-\Omega^2_+}\right]=\chi_{yy}(\omega)\\
		\chi_{xy}(\omega)&=&\frac{2J^2_0}{\hbar}
		\left[\frac{(\cosh r -\zeta \sinh r)^2 i(\omega+i\gamma)}{(\omega+i\gamma)^2-\Omega^2_-}-\frac{(\sinh r -\zeta \cosh r)^2 i (\omega+i \gamma)}{(\omega+i \gamma)^2-\Omega^2_+}\right]\end{eqnarray}
	
	Similarly, for the CRA case we have
	\begin{eqnarray}
		\chi_{xx}(\omega)&=&\frac{2J^2_0}{\hbar}\left[\frac{\left(\cos \theta -\zeta \sin \theta\right)^2\Omega_-}{(\omega+i\gamma)^2-\Omega^2_-}+\frac{(\sin \theta +\zeta \cos \theta)^2 \Omega_+}{(\omega+\textrm{i}\gamma)^2-\Omega^2_+}\right]=\chi_{yy}(\omega)\\
		\chi_{xy}(\omega)&=&\frac{2J^2_0}{\hbar}
		\left[\frac{(\cos \theta -\zeta \sin \theta)^2 i(\omega+i\gamma)}{(\omega+i\gamma)^2-\Omega^2_-}+\frac{(\sin \theta +\zeta \cos \theta)^2 i (\omega+i \gamma)}{(\omega+i \gamma)^2-\Omega^2_+}\right]\end{eqnarray}.

	\subsection{DC Conductivities}
	\label{sec:dc conductivity}
	
	Using the Kubo formula~\cite{kuboStatisticalMechanicalTheoryIrreversible1957}, the optical conductivity tensor is given by
	\begin{equation}\label{Eq: kubo formula}
		\sigma_{ab}(\omega) = \frac{i}{\omega + i\gamma}\left[\frac{e^2 n_{\mathrm{2D}}}{m^*}\delta_{ab} + \frac{\chi_{ab}(\omega)}{A}\right],
	\end{equation}
	where $n_{\mathrm{2D}} = N/A$ is the 2D electron density. The first term is the Drude (diamagnetic) contribution; the second is the paramagnetic (current-current) correction.
	
	Setting $\omega = 0$ for the DC conductivity, we obtain 
	
	\paragraph{Longitudinal conductivity.} 
	\begin{equation}
		\sigma_{xx}(0) = \frac{1}{\gamma}\left[\frac{e^2 n_{\mathrm{2D}}}{m^*} - \frac{J_0^2}{A\hbar}\sum_r \frac{2\Omega_r\,|\Lambda_{x,r}|^2}{\Omega_r^2 + \gamma^2}\right].
	\end{equation}
	Setting the Drude conductivity $\sigma_D = e^2 n_{\mathrm{2D}}/(m^*\gamma)$, we can rewrite
	\begin{equation}\label{Eq: sigma xx dc}
		\frac{\sigma_{xx}^{\mathrm{dc}}}{\sigma_D} = 1 - \sum_r\frac{\omega_{\mathrm{cyc}}\,\Omega_r\,|\Lambda_{x,r}|^2}{\Omega_r^2 + \gamma^2}.
	\end{equation}
	
	\paragraph{Hall conductivity.}
	\begin{equation}
		\sigma_{xy}(0) = \frac{1}{\gamma}\cdot\frac{e^2 n_{\mathrm{2D}}\omega_{\mathrm{cyc}}}{2m^*}\sum_r \Lambda_{x,r}\Lambda_{y,r}\left[\frac{1}{-\Omega_r + i\gamma} + \frac{1}{\Omega_r + i\gamma}\right].
	\end{equation}
	Using $e^2 n_{\mathrm{2D}}\omega_{\mathrm{cyc}}/m^* = (e^2\nu/h)\omega_{\mathrm{cyc}}^2$ where $\nu = n_{\mathrm{2D}}h/(eB)$:
	\begin{equation}\label{Eq: sigma xy dc}
		\frac{\sigma_{xy}^{\mathrm{dc}}}{\nu e^2/h} = \sum_r\frac{\omega_{\mathrm{cyc}}^2\,\Lambda_{x,r}\Lambda_{y,r}}{\Omega_r^2 + \gamma^2}.
	\end{equation}
	
	Following Ref.~\cite{BartoloCiuti2018}, the Shubnikov--de Haas oscillations enter through a field-dependent scattering rate,$\gamma_\mathrm{r}(B) = \frac{W_{\mathrm{e}, r}}{\tau_e}  + \frac{W_{\rm p, r}}{\tau_p}$. The photonic and electronic weights $W_{\mathrm{e}, r}$ and $W_{\mathrm{p}, r}$ are obtained by Eq.~\ref{Eq: Wer Wpr}. Here, the electron transport time is given by, $\frac{1}{\tau_e} = \frac{1}{\tau_0}\left[1 - 2\exp\left(-\frac{\pi}{\tau_\mathrm{q}\, \omega_{\mathrm{cyc}}}\right) \cos(2\pi \nu)\right]$. For the simulations, we assume $\tau_0\, \omega_\mathrm{cav} = 100 $, $\tau_\mathrm{q}\, \omega_\mathrm{cav} = 2$, and $\tau_\mathrm{p}\, \omega_\mathrm{cav} = 1000.$ The DC transport quantities now include field-dependent $\gamma_\mathrm{r}(B)$.
	
	\subsection{Uncoupled limit and sum rules}
	\label{sec:uncoupled}
	
	At $g = 0$: one polariton branch at $\Omega = \omega_{\mathrm{cyc}}$ with $\Lambda_x = \Lambda_y = 1$. The DC formulas reduce to the standard Drude magnetoconductivity:
	\begin{equation}\label{Eq: drude}
		\frac{\sigma_{xx}^{\mathrm{dc}}}{\sigma_D} = \frac{\gamma^2}{\omega_{\mathrm{cyc}}^2 + \gamma^2}, \qquad \frac{\sigma_{xy}^{\mathrm{dc}}}{\nu e^2/h} = \frac{\omega_{\mathrm{cyc}}^2}{\omega_{\mathrm{cyc}}^2 + \gamma^2}.
	\end{equation}
	In the disorder-less limit $\gamma \to 0$, we should obtain $\sigma_{xx} \to 0$ and $\sigma_{xy} \to \nu e^2/h$ to recover quantization. Therefore, for the cavity-coupled system to preserve this limit, the Bogoliubov coefficients satisfy:
	\begin{equation}  
		\label{Eq: sum rules}
		\sum_r \frac{\omega_{\mathrm{cyc}}\,|\Lambda_{x,r}|^2}{\Omega_r} = 1,\qquad\sum_r \frac{\omega_{\mathrm{cyc}}^2\,\Lambda_{x,r}\Lambda_{y,r}}{\Omega_r^2} = 1. 
	\end{equation}
	These sum rules guarantee that the disorder-less values $\sigma_{xx} = 0$ and $\sigma_{xy} = \nu e^2/h$ are exact.
	
	\subsection{DC Conductivity $\gamma$ expansion}
	
	Expanding the denominators at small $\gamma$:
	\begin{equation*}
		\frac{1}{\Omega_r^2 + \gamma^2} = \frac{1}{\Omega_r^2} - \frac{\gamma^2}{\Omega_r^4} + O(\gamma^4).
	\end{equation*}
	The DC conductivities become
	\begin{align}
		\frac{\sigma_{xx}^{\mathrm{dc}}}{\sigma_D} &= \left(1 - \omega_{\mathrm{cyc}}\sum_r \frac{|\Lambda_{x,r}|^2}{\Omega_r}\right) + \gamma^2\,\omega_{\mathrm{cyc}}\sum_r \frac{|\Lambda_{x,r}|^2}{\Omega_r^3} + O(\gamma^4) =  \gamma^2\,\omega_{\mathrm{cyc}}\sum_r \frac{|\Lambda_{x,r}|^2}{\Omega_r^3} + O(\gamma^4)\\
		\frac{\sigma_{xy}^{\mathrm{dc}}}{\nu e^2/h} &= \omega_{\mathrm{cyc}}^2\sum_r \frac{\Lambda_{x,r}\Lambda_{y,r}}{\Omega_r^2} - \gamma^2\omega_{\mathrm{cyc}}^2\sum_r \frac{\Lambda_{x,r}\Lambda_{y,r}}{\Omega_r^4} + O(\gamma^4)= 1 - \gamma^2\omega_{\mathrm{cyc}}^2\sum_r \frac{\Lambda_{x,r}\Lambda_{y,r}}{\Omega_r^4} + O(\gamma^4),
	\end{align}
	where we have used the sum rules in Eq.~\ref{Eq: sum rules}.
	
	The CRI Matrix in Eq.~\ref{Eq: m cri}, decouples into two 2$\times$2 blocks. 
	From Eq.~\ref{Eq: m cri}, we have:
	\begin{equation}
		y_- = \frac{\Omega_- - \omega_{\mathrm{cyc}}}{\bar g}\,x_-, \qquad z_+ = \frac{\Omega_+ - \Omega}{\bar g}\,w_+, 
		\label{Eq: wxyz forms}
	\end{equation}
	with $w_- = z_- = 0$ and $x_+ = y_+ = 0$.  
	
	From the eigenvalues of Eq.~\ref{Eq: m cri}, we find the relations
	\begin{equation}
		\Omega_- \Omega_+ = \omega_{\mathrm{cav}} \omega_{\mathrm{cyc}}, \qquad \bar g^2 =  (\omega_{\mathrm{cyc}} - \Omega_-)(\Omega + \Omega_-) = (\Omega - \Omega_+)(\omega_{\mathrm{cyc}} + \Omega_+).
	\end{equation}
	
	Using the Bogoliubov normalization condition, $w_r^2 + x_r^2 - y_r^2 - z_r^2 = 1$, and the above identities, we can obtain
	\begin{equation}
		x_-^2 = \frac{\bar g^2}{(\omega_{\mathrm{cyc}} - \Omega_-)(\Omega + 2\Omega_- - \omega_{\mathrm{cyc}})}, \qquad w_+^2 = \frac{\bar g^2}{(\Omega - \Omega_+)(\omega_{\mathrm{cyc}} 	+ 2\Omega_+ - \Omega)}
		\label{Eq: x2 and w2}
	\end{equation}
	
	The current weights in Eq.~\ref{Eq: currentweight} reduce to 
	\begin{align*}
		\Lambda_{x, -} = x_- + \zeta y_-,&\quad  \Lambda_{x, +} = -(z_+ + \zeta w_+), \\
		\Lambda_{y, -} = \Lambda_{x, -} ,&\quad  \Lambda_{y, +} = -(\Lambda_{x, +}).
	\end{align*}
	rewriting using Eq.~\ref{Eq: x2 and w2}, 
	\begin{equation}
		|\Lambda_{x,+}|^2 = \frac{(\Omega_+ - \omega_{\mathrm{cav}})^2}{(\Omega - \Omega_+)(\omega_{\mathrm{cyc}} + 2\Omega_+ - \Omega)}, \qquad |\Lambda_{x,-}|^2 = \frac{\Omega_-^2\,\bar g^2}{\omega_{\mathrm{cyc}}^2\,(\omega_{\mathrm{cyc}} - \Omega_-)(\Omega + 2\Omega_- - \omega_{\mathrm{cyc}})}.
	\end{equation}
	Substituting the above equations, and identifying $\Lambda_{x,+}\Lambda_{y,+} = -|\Lambda_{x,+}|^2$ and $\Lambda_{x,-}\Lambda_{y,-} = +|\Lambda_{x,-}|^2$, we obtain:
	\begin{equation}
		\omega_{\mathrm{cyc}}\sum_r \frac{|\Lambda_{x,r}|^2}{\Omega_r^3} = \frac{\Omega}{\omega_{\mathrm{cav}}\,\omega_{\mathrm{cyc}}^2}, \quad
		\omega_{\mathrm{cyc}}^2\sum_r \frac{\Lambda_{x,r}\Lambda_{y,r}}{\Omega_r^4} = \frac{1}{\omega_{\mathrm{cyc}}^2} - \frac{D(\omega_{\mathrm{cyc}} - 2\omega_{\mathrm{cav}} - D)}{\omega_{\mathrm{cav}}^2\,\omega_{\mathrm{cyc}}^2}.
	\end{equation}
	
	The CRA case is obtained in the same manner, 
	\begin{equation*}
		\omega_{\mathrm{cyc}}^2\sum_r \frac{\Lambda_{x,r}\Lambda_{y,r}}{\Omega_r^4} = \frac{1}{\omega_{\mathrm{cyc}}^2} + \frac{D(\omega_{\mathrm{cyc}} + 2\omega_{\mathrm{cav}} + D)}{\omega_{\mathrm{cav}}^2\,\omega_{\mathrm{cyc}}^2}.
	\end{equation*}
	
	\subsection{DC Conductivities: Closed-form corrections}
	
	Substituting back into the $\gamma$ expansion and dividing by the bare values (From Eq.~\ref{Eq: drude}, $\sigma_{xx}^{\mathrm{bare}}/\sigma_D = \gamma^2/(\omega_{\mathrm{cyc}}^2 + \gamma^2)$ and $\sigma_{xy}^{\mathrm{bare}}/(\nu e^2/h) = \omega_{\mathrm{cyc}}^2/(\omega_{\mathrm{cyc}}^2 + \gamma^2)$). At leading order in $\gamma$, we have $\sigma_{xx}^{\mathrm{bare}}/\sigma_D = \gamma^2/\omega_{\mathrm{cyc}}^2$ and $\sigma_{xy}^{\mathrm{bare}}/(\nu e^2/h) = 1.$ 
	\begin{equation}
		\frac{\Delta\sigma_{xx}}{\sigma_{xx}^{\mathrm{bare}}} = \frac{D}{\omega_{\mathrm{cav}}} = \frac{g^2}{\omega_{\mathrm{cav}}^2}, \qquad \frac{\Delta\sigma_{xy}^{(s)}}{\sigma_{xy}^{\mathrm{bare}}} = \frac{\gamma^2\,D\,(s\,\omega_{\mathrm{cyc}} - 2\omega_{\mathrm{cav}} - D)}{\omega_{\mathrm{cav}}^2\,\omega_{\mathrm{cyc}}^2}.
	\end{equation}
	
	with $s = 1$ (CRI) and $s = -1$ (CRA).

	For the achiral case, the current weights $\Lambda^\mathrm{ach}_{x/y, r}$ and the Hopfield matrix in Eq.~\ref{Eq: m linear}, give
	\begin{equation}
		\label{Eq: achiral conductivites}
		\frac{\Delta\sigma_{xx}^{\mathrm{ach}}}{\sigma_{xx}^{\mathrm{bare}}} = 	\frac{2D}{\omega_{\mathrm{cav}}} = \frac{2g^2}{\omega_{\mathrm{cav}}^2}, \qquad \frac{\Delta\sigma_{xy}^{\mathrm{ach}}}{\sigma_{xy}^{\mathrm{bare}}} = -\frac{4\gamma^2\,D\,\Omega}{\omega_{\mathrm{cav}}^2\,\omega_{\mathrm{cyc}}^2}.
	\end{equation}
	
	\section{Resistivity tensor}
	\label{sec:resistivity}

	We invert the DC conductivity tensor, $\rho_{xx} = \sigma_{xx}/(\sigma_{xx}^2 + \sigma_{xy}^2)$ and $\rho_{xy} = \sigma_{xy}/(\sigma_{xx}^2 + \sigma_{xy}^2)$, starting from the leading-order conductivities of the previous section. With the sum rules $S_1 = S_2 = 1$ and the evaluated sums $S_3$ and $S_4$, the conductivities read
	\begin{equation}
		\sigma_{xx} = \frac{\nu e^2}{h}\,\frac{\gamma\,\Omega}{\omega_{\mathrm{cav}}\,\omega_{\mathrm{cyc}}},
		\qquad
		\sigma_{xy} = \frac{\nu e^2}{h}\left[1 - \frac{\gamma^2\,D\,(s\,\omega_{\mathrm{cyc}} - 2\omega_{\mathrm{cav}} - D)}{\omega_{\mathrm{cav}}^2\,\omega_{\mathrm{cyc}}^2}\right],
		\label{Eq: sigma for inversion}
	\end{equation}
	with $s = +1$ (CRI) and $s = -1$ (CRA).
	
	The determinant (up to order $\gamma^2$), is
	\begin{equation}
		\sigma_{xx}\sigma_{yy} - \sigma_{xy}\sigma_{yx} = \sigma_{xx}^2 + \sigma_{xy}^2
		= \left(\frac{\nu e^2}{h}\right)^2\left[\frac{\gamma^2\,\Omega^2}{(\omega_{\mathrm{cav}}\omega_{\mathrm{cyc}})^2}
		+ 1- \frac{\gamma^2\,D\,(s\,\omega_{\mathrm{cyc}} - 2\omega_{\mathrm{cav}} - D)}{\omega_{\mathrm{cav}}^2\,\omega_{\mathrm{cyc}}^2}\right]
		= \left(\frac{\nu e^2}{h}\right)^2\left[1 - \frac{\gamma^2(\Omega^2 - 2s\bar g^2)}{(\omega_{\mathrm{cav}}\omega_{\mathrm{cyc}})^2}\right].
		\label{Eq: determinant}
	\end{equation}
	Dividing Eq.~\eqref{Eq: sigma for inversion} by Eq.~\eqref{Eq: determinant}, only the handedness-dependent term survives:
	\begin{equation}
		\rho_{xy}^{(s)} = \frac{h}{\nu e^2}\left[1 - \frac{s\,\gamma^2\,\bar g^2}{(\omega_{\mathrm{cav}}\,\omega_{\mathrm{cyc}})^2}\right]
		= \frac{h}{\nu e^2} - \frac{s\,m^*\gamma^2 D}{n_{\mathrm{2D}}e^2\,\omega_{\mathrm{cav}}^2},
		\label{Eq: rho xy chiral}
	\end{equation}
	where we used $\bar g^2 = D\,\omega_{\mathrm{cyc}}$ and $h/(\nu e^2) = m^*\omega_{\mathrm{cyc}}/(n_{\mathrm{2D}}e^2)$. The crossover at $\omega_{\mathrm{cyc}}^* = 2\omega_{\mathrm{cav}} + D$ present in $\Delta\sigma_{xy}$ does not show up in $\rho_{xy}$. The longitudinal resistivity is
	\begin{equation}
		\rho_{xx} = \frac{\sigma_{xx}}{\sigma_{xx}^2 + \sigma_{xy}^2}
		= \frac{m^*\gamma}{n_{\mathrm{2D}}e^2}\left(1 + \frac{g^2}{\omega_{\mathrm{cav}}^2}\right),
		\label{Eq: rho xx chiral}
	\end{equation}
	identical for both handednesses.
	
	For the achiral cavity, using Eq.~\ref{Eq: achiral conductivites} and repeating the inversion, we have,
	\begin{equation}
		\sigma_{xx}^2 + \sigma_{xy}^2 = \left(\frac{\nu e^2}{h}\right)^2\left[1 - \frac{\gamma^2(\omega_{\mathrm{cav}}+2D)^2}{(\omega_{\mathrm{cav}}\omega_{\mathrm{cyc}})^2}\right]
		\quad\Longrightarrow\quad
		\rho_{xy}^{\mathrm{ach}} = \frac{h}{\nu e^2},
		\label{Eq: rho xy achiral}
	\end{equation}
	and
	\begin{equation}
		\rho_{xx}^{\mathrm{ach}} = \frac{m^*\gamma}{n_{\mathrm{2D}}e^2}\left(1 + \frac{2g^2}{\omega_{\mathrm{cav}}^2}\right).
		\label{Eq: rho xx achiral}
	\end{equation}

	\section{Normalized current quadrature}
	Defining $\hat{c} = ( \hat{b} - \zeta\,\hat{a})/\sqrt{1+\zeta^2}$, which satisfies $[\hat{c}, \hat{c}^\dagger] = 1$, the y-component of the current, Eq.~\ref{Eq: jxjy} becomes
	\begin{equation}
		{\hat J}_y = -iJ_0 \sqrt{1 + \zeta^2} \left(\hat c - \hat c^\dagger\right); \quad {\hat j}_y =  -i\left(\hat c - \hat c^\dagger\right)/2
	\end{equation}
	where ${\hat j}_y$ is the normalized current quadrature operator.  Variance of an operator $\hat{\mathcal{O}}$ is given by
	\begin{equation*}
		\langle(\Delta\hat{\mathcal{O}})^2\rangle 
		= \langle G|\hat{\mathcal{O}}^2|G\rangle 
		- \langle G|\hat{\mathcal{O}}|G\rangle^2.
	\end{equation*}
	Substituting the above equation to calculate the variance of the normalized current quadrature and noting that $\langle G|{\hat j}_y|G\rangle = 0$, we have
	\begin{equation}
		\langle (\Delta{\hat j}_y)^2\rangle = \langle G|{\hat j}_y^2|G\rangle=- \frac{1}{4(1+\zeta^2)}\langle G|
		\bigl[(\hat{b}-\hat{b}^\dagger) + s \zeta(\hat{a}-\hat{a}^\dagger)\bigr]^2|G\rangle.
	\end{equation}
	Substituting the inverse Bogoliubov transformation~(\ref{Eq: inverse ab}), the variance is therefore
	\begin{equation}\label{Eq: variance}
		\langle (\Delta{\hat j}_y)^2\rangle
		= \frac{\sum_r|\Lambda_{y,r}|^2}{4(1+\zeta^2)}.
	\end{equation}
	Analogous expressions hold for ${\hat j}_x$ with 
	$\Lambda_{x,r} = (x_r - z_r) - \,\zeta(w_r - y_r)$.
	
	\paragraph{CRA:}
	
	For the CRA Hamiltonian, $y_r = z_r = 0$, so 
	$\Lambda_{y,r} = x_r - \zeta\,w_r$. The eigenvectors satisfy 
	$\sum_r w_r^2 = 1$, $\sum_r x_r^2 = 1$, $\sum_r w_r x_r = 0$. Therefore 
	$\sum_r|\Lambda_{x,r}|^2 = 1 + \zeta^2$ and 
	$\langle(\Delta{\hat j}_y)^2\rangle = 1/4$ exactly.
	
	\paragraph{Achiral case:}
	
	For the achiral Hamiltonian, the current contains contributions from both 
	photon modes. Defining  $\hat{c} = (\hat{b} - \zeta\,\hat{a}_R - \zeta\,\hat{a}_L)/\sqrt{1+2\zeta^2}$, the current quadrature variance takes the form
	\begin{equation}
		\langle (\Delta{\hat j}_y)^2\rangle
		= \frac{\sum_r|\Lambda_{y,r}^{\mathrm{ach}}|^2}{4(1+2\zeta^2)},
	\end{equation}
	where $\Lambda_{y,r}^{\mathrm{ach}}$ are the achiral current weights, and the sum runs over the three 
	positive-frequency polariton branches obtained from the $6\times6$ Hopfield matrix (Eq.~\ref{Eq: m linear}). Both CRI and achiral variances are presented 
	in Fig.~2 of the main text.
	
	\section{Coupling strength for a 1D photonic-crystal cavity}
	\label{sec:g estimate}
	
	The collective coupling $g = \sqrt{Ne^2/(2\epsilon_r\epsilon_0\mathcal{V}m^*)}$ depends on the cavity geometry through the effective mode volume $\mathcal{V}$. For a one-dimensional photonic-crystal cavity with the 2DEG located at $z_{\mathrm{2DEG}}$, $\mathcal{V}$ is determined by the cavity mode profile, following the procedure of Ref.~\cite{liVacuumBlochSiegert2018}. The mode profile $f(z) \propto \sqrt{\epsilon_{\mathrm{cav}}(z)}\,\mathcal{E}(z)$ is obtained from a transfer-matrix calculation of the empty cavity at its resonance $\omega_{\mathrm{cav}}$, normalized over the cavity interior, $\int dz\,|f(z)|^2 = 1$. Due to field confinement along one direction, we define effective mode length
	\begin{equation}
		L_{\mathrm{eff}} = \frac{2}{|f(z_{\mathrm{2DEG}})|^2}.
	\end{equation}
	We, therefore, have the light-matter coupling at zero detuning 
	\begin{equation}
		g = \sqrt{\frac{n_{\mathrm{2D}}\,e^2}{\epsilon_r\epsilon_0\,m^*\,L_{\mathrm{eff}}}},
		\label{Eq: g pcc}
	\end{equation}
	
	For our parameters ($n_{\mathrm{2D}} = 5\times10^{11}\,\mathrm{cm}^{-2}$ per well for $10$ quantum wells, $m^* = 0.067\,m_e$, $\epsilon_r = 12.96$), transfer-matrix simulation for a chiral cavity designed for $\omega_\mathrm{cav}/2\pi=0.35$~THz yields a light--matter coupling strength of $g/\omega_\mathrm{cav}\sim0.42$.

	\bibliography{references}